# Machine Learning of Partial Differential Equations from Noise Data

Wenbo Cao[1], Weiwei Zhang[2*]

1 School of aeronautics, Northwestern Polytechnical University, Xi'an 710072, China. 308967347@qq.com. ORCID: 0000-0002-5991-1816.

2* School of aeronautics, Northwestern Polytechnical University, Xi'an 710072, China. E-mail: aeroelastic@nwpu.edu.cn. ORCID: 0000-0001-7799-833X. (Corresponding author).

**Abstract** Machine learning of partial differential equations from data is a potential breakthrough to solve the lack of physical equations in complex dynamic systems, but noise has become the biggest obstacle in the application of partial differential equation identification method, since numerical differentiation is ill-posed to noise data. To overcome this problem, we propose Frequency Domain Identification method based on Fourier transform. This method first conducts Fourier transform on each item in the candidate library, and then uses their low-frequency component to identify partial differential equations, which ensures that the data still obey the original partial differential equation after filtering the noise, so it has a good partial differential equations identification effect for strong noise data. We also propose a new sparse identification criterion, which can accurately identify the terms in the equation from low signal-to-noise ratio data. Through identifying a variety of canonical equations spanning a number of scientific domains, the proposed method is proved to have high accuracy and robustness for equation structure and parameters identification for low signal-to-noise ratio data. The method provides a promising technique to discover potential partial differential equations from noisy experimental data.

## 1 Introduction

Partial differential equations (PDEs) play a more and more important role in modern science. It is used to describe mathematical laws behind physical systems and plays a vital role in the analysis, prediction and control of many systems. In the past, PDEs were derived mathematically or physically through basic conservation laws, which resulted in many canonical models in physics, engineering and other fields, such as Navier-Stokes equations describing fluid motion, Maxwell's equations describing electromagnetic fields and so on, which greatly promoted the progress of science. However, in modern applications, the mechanism of many complex systems is still unclear, and it is difficult to derive PDEs (for example, multiphase flow, neuroscience, finance, bioscience and so on). In the past decade, with the rapid development of sensors, computing power and data storage, the cost of data collection and computing has been greatly reduced so that we can obtain a large amount of experimental data, and the rapid development of machine learning [1] has also provided a reliable tool to discover the potential laws of the system from large datasets. Nowadays, the machine learning of differential equations has become a promising new technology to discover physical laws in complex systems.

Advanced regression methods in machine learning, such as genetic programming or sparse regression, are driving new techniques that identify differential equations from measurements of complex systems. The symbolic regression method [2,3] first obtain the ordinary differential equations from data by genetic programming [4], and it realizes the long-sought goal of identifying physical laws from data. However, its calculation is slow, and prone to overfitting. The sparse identification of nonlinear dynamics method (SINDy) [5] is a recent breakthrough which uses the sparse regression and a Pareto analysis to discover differential equations from many potential dynamical models. The only assumption about the structure of the model in this method is that there are only a few important terms that govern the dynamics, so that the equations are sparse in the possible function space. SINDy has been widely used and developed in many fields, such as discovery of disease dynamics [6], discovery of vortex-induced vibration [7], rapid model recovery from abrupt system changes [8], discovery of algebraic Reynolds-Stress model [9], discovery of subsurface flow equations [10], discovery of multiscale nonlinear dynamics [11], model selection via sparse regression and Akaike information criteria [12], Galerkin regression models [13] and so on.

PDE functional identification of nonlinear dynamics (PDE-FIND) [14] develops the SINDy to identify PDEs from parameterized spatiotemporal data, which selects the terms that most accurately represent the data from an over-completed library containing many nonlinear and partial derivative terms by the sparse regression. PDE-FIND is so simple and effective that the potential PDE of the data can be obtained directly. However, it is difficult to identify PDEs from low signal-to-noise ratio (SNR) data [15] because the numerical differentiation is ill-posed to noise data [16]. A similar method is also independently proposed in [17], which adds noise to the time derivative after it was computed from clean data instead of directly adding noise to the original data. PDE-net [18,19] introduces the discretization of differential equations into convolution neural networks through the relationship between differential operators and convolution kernels [20]. PDE-net can accurately predict dynamics of complex systems and get the PDE of the data, but the paper only shows the effect of this method on the data with very small noise. Data-driven method and data-assimilation method are combined [21] to identify partial differential equations and its additional model parameters from data, which broadens the applicable area of PDE identification methods. When the structure of PDEs is known, the Gaussian process [22,23] is used to identify the parameters of the PDE, which requires less data than the sparse regression. A physical information neural network (PINN) [24] is proposed to identify the parameters of PDEs. This method uses neural networks to fit the data, and uses automatic differentiation [25] to take PDE as a regularization term in the fitting process, which avoids the ill-posed problem of traditional numerical differentiation method. In the application of PDEs identification methods, PDE-FIND is used to identify various hydrodynamic equations from the simulated flow field obtained by molecular simulation [26], which shows the correlation between Boltzmann equation and Navier-Stokes equations. It is also mentioned in the paper that PDE-FIND is only suitable for high signal-to-noise ratio data.

To solve the problem that existing PDE identification methods are ineffective to low SNR data, we propose Frequency Domain Identification method. In this method, we do not deal with the noise of $u(x,t)$ directly, because any change of $u$ may cause it to no longer satisfy the original equation. On the contrary, we first calculate each item in the candidate library, and then conduct the Fourier transform to it. Finally, we use its low-frequency components to identify the equation, this method ensures that the data still obey the original PDE while filtering the noise, so it has a good effect for strong noise data. This paper also proposes a new sparse recognition criterion to identify the structure of partial differential equations, which can identify terms with small coefficients in partial differential equations.

## 2  Materials and methods

In this part, we first briefly introduce PDE-FIND [14], and then propose Frequency Domain Identification and a new sparse identification method.

2.1 PDE-FIND

PDE-FIND first builds a library containing many of partial differential terms and then selects the terms that best match the data by the sparse regression. The specific process is as follows.

Consider the general form of nonlinear PDE: $u_t = N(u, u_x, u_{xx}, \cdots, x, \mu)$, where the subscripts denote partial differentiation in either time or space, and N(·) is an unknown nonlinear function. The method builds an over-completed library that contains all the terms may appear in the PDE, such as:

$$\varphi = [u, u^2, u_x, uu_x, u^2 u_x, u_{xx}, uu_{xx}, u^2 u_{xx}, u_{xxx}, uu_{xxx}, u^2 u_{xxx}] \tag{1}$$

In this case, PDE can be expressed as a linear combination of each term in the library, that is,

$$u_t = \varphi \xi \tag{2}$$

where $\varphi$ represents the library, and vector $\xi$ represents each term of the coefficients in the library corresponding to PDE. The vector $\xi$ uniquely determines PDE, as an example, the Burgers' equation $u_t + uu_x - 0.03u_{xx} = 0$ can be expressed as $u_t = \varphi[0,0,0,-1,0,0.03,0,0,0,0,0]^T$, so the PDE of the system can be obtained only by calculating the vector $\xi$ from the experimental data.

For the given experimental data of a physical field, the values of $u_t$ and $\varphi$ at many time-space positions are calculated and arranged as column vectors respectively, and Eq. 3 is obtained. Since every point in the field should satisfy $u_t = \varphi \xi$, the PDE of the field can be obtained by solving $\xi$. A key assumption is that the PDE consists of only a few terms, so that the vector $\xi$ is sparse. To obtain the sparse vector $\xi$, PDE-FIND uses the Sequential Threshold Ridge regression to solve the matrix equation.

$$\underbrace{\begin{pmatrix} u_t(x_1,t_1) \\ u_t(x_2,t_2) \\ \vdots \\ u_t(x_m,t_m) \end{pmatrix}}_{U_t} = \underbrace{\begin{pmatrix} u(x_1,t_1) & u^2(x_1,t_1) & \cdots & u^2 u_{xxx}(x_1,t_1) \\ u(x_2,t_2) & u^2(x_2,t_2) & \cdots & u^2 u_{xxx}(x_2,t_2) \\ \vdots & \vdots & & \vdots \\ u(x_m,t_m) & u^2(x_m,t_m) & \cdots & u^2 u_{xxx}(x_m,t_m) \end{pmatrix}}_{\Phi} \xi \qquad (3)$$

2.2 Frequency Domain Identification

PDE-FIND is simple and effective, but it is difficult to identify PDE from low SNR data [15] because the numerical differentiation is ill-posed to noise data [16]. In this paper, we propose Frequency Domain Identification to solve the thorny problem.

As mentioned in Section 2.1, any linear or nonlinear partial differential equation can be expressed as a linear combination of each term in the library (Eq. 2). Therefore, when the same linear transformation is carried out on $u_t$ and $\varphi$, the equation is still hold, that is,

$$L(u_t) = L(\varphi)\xi \qquad (4)$$

Where $L$ is a linear transformation, $L(\varphi)$ denotes $[L(\varphi_1), L(\varphi_2), \cdots, L(\varphi_n)]$. Obviously, Eq. 4 is the same as $\xi$ in Eq. 2. When $\xi$ is obtained by Eq. 4, we can get the partial differential equation which accords with the data. In this case, if $L$ is a linear transformation to filter noise, the effect of filtering noise can be achieved, and the transformation $L$ will have not any adverse effect on $\xi$, because for the linear transformation $L$, Eq. 4 holds accurately. It should be noted that $L$ can be different in different space-time positions, because we only need to ensure that Eq. 4 holds in any space-time positions, and then we can rearrange them into equations similar to Eq. 3 to solve the vector $\xi$. This means that any linear method of filtering noise that can be "$L$". This is very important, because if we do a linear transformation on $u(x,t)$ to filter noise according to the general idea, when there is a nonlinear term in the PDE, $L(u)$ no longer satisfies with the original PDE. On the other hand, if we do linear transformation on $u_t$ and $\varphi$ to filter noise, Eq. 4 still holds so that we can get $\xi$.

As mentioned above, any linear noise filtering method can be applied to this framework. In this paper, we draw lessons from the low-pass filtering. The low pass filtering transforms the noise signal to frequency domain, selects a certain frequency threshold, and applies inverse Fourier transform to the spectrum which is less than the threshold, which can well filter out the high frequency noise in the signal. However, there are two problems in the direct low pass filtering to noise signal. 1. The noise cannot be completely filtered. In Fig. 1, the frequency spectrums of the noise and signal overlap between threshold 1 and threshold 2, and the low pass filtering using threshold 1 will cause signal distortion, while using threshold 2 cannot completely eliminate noise. Neither of them can get good results, and the low pass filtering can only make a tradeoff between threshold 1 and threshold 2. 2. It is difficult to select the frequency domain threshold. For the experimental data, there is no true value as a control. It is difficult to distinguish the spectrums of the signal and noise to select the threshold.

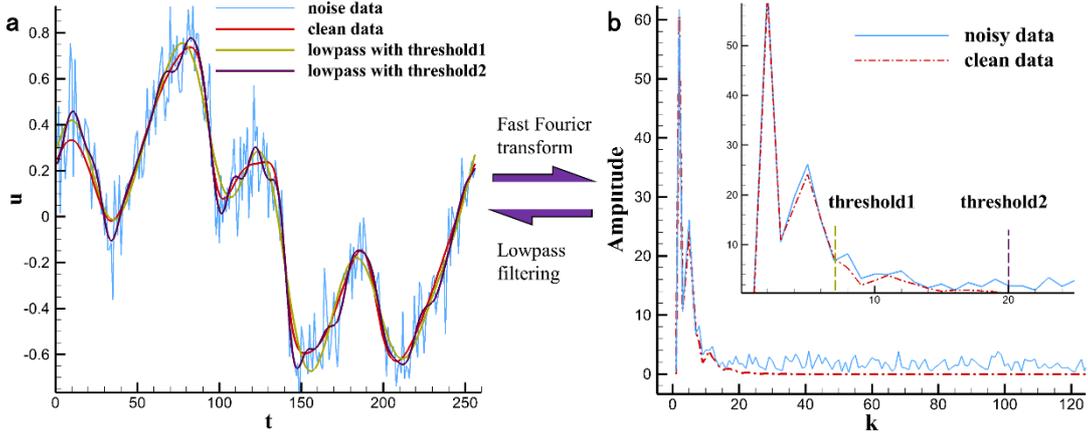

**Fig. 1.** Effects of low pass filter. (a) Original noise and noiseless signals, and the signals after the low pass filtering with threshold 1 and threshold 2. (b) The spectrums of noise and noiseless signals.

So, we directly conduct the Fourier transform on $u_t$ and $\varphi$, since the Fourier transform is a linear operation, every point in the frequency domain is still accurately satisfied with Eq. 4. Therefore, we can choose any small cutoff frequency without having to worry about discarding part of the real data, which can filter the noise to the maximum extent. At the same time, we can also directly use the frequency domain data to solve $\xi$, without the inverse Fourier transform on the filtered data, of course, the results of the two methods are the same. Moreover, a very important point is that we can conduct Fourier transform to each dimension of the data, for example, it can be performed about x and t successively for one-dimensional PDE (Eq. 5), so that the low frequency data is not only low frequency in time but also in all dimensions of space, which can make the low frequency component have as little noise as possible. It should be noted that we do not call this process low-pass filtering but named it Frequency Domain Identification, because in general, the purpose of low-pass filtering is to recover the real signal as much as possible, but for the Frequency Domain Identification, in order to filter the noise as much as possible, we can filter out part of the real signal at the same time, because the rest of the signal is always satisfied with Eq. 4.

$$\begin{cases} u_t(x,t) = \xi_1 u u_x(x,t) + \xi_2 u_{xx}(x,t) \\ \mathcal{F}_t[u_t(x,k_t)] = \xi_1 \mathcal{F}_t[u u_x(x,k_t)] + \xi_2 \mathcal{F}_t[u_{xx}(x,k_t)] \\ \mathcal{F}_t\mathcal{F}_x[u_t(k_x,k_t)] = \xi_1 \mathcal{F}_t\mathcal{F}_x[u u_x(k_x,k_t)] + \xi_2 \mathcal{F}_t\mathcal{F}_x[u_{xx}(k_x,k_t)] \end{cases} \quad (5)$$

It is worth noting that the natural way to calculate the frequency domain value of the derivative term in the library is through the equation $\mathcal{F}\left[\frac{\partial^n u(x)}{\partial x^n}\right] = (jk)^n \mathcal{F}[u(x)]$. However, we do not use this method to calculate, because $u$ is required to be periodic when calculating the partial differential for finite length discrete data, which will limit the application of Frequency Domain Identification. Therefore, we first calculate the time-space domain value of each term, and then get its two-dimensional Fourier transform, which avoids the periodic restriction of the data.

Fig. 2 shows the process of Frequency Domain Identification in detail. Compared with PDE-FIND, after obtaining the function value matrix $\Phi_i$ corresponding to each term in $\varphi$, we apply two-dimensional Fourier transform to each matrix $\Phi_i$ to get $\widehat{\Phi_i}$ (for a high-dimensional PDE, we can carry out three-dimensional or four-dimensional Fourier transform). Then the low frequency component of the frequency domain data (the red region in Fig. 2) is selected to form the frequency domain matrix equation:

$$\widehat{U_t} = \widehat{\Phi}\xi \quad (6)$$

In Eq. 6, $\widehat{U_t}$ and $\widehat{\Phi}$ represent the results of two-dimensional Fourier transform of $U_t$ and $\Phi$. By solving the frequency domain matrix equation, we can get the frequency domain form of the PDE. We can also get the time-space domain form of the PDE because the frequency domain form corresponds to the time-space domain form.

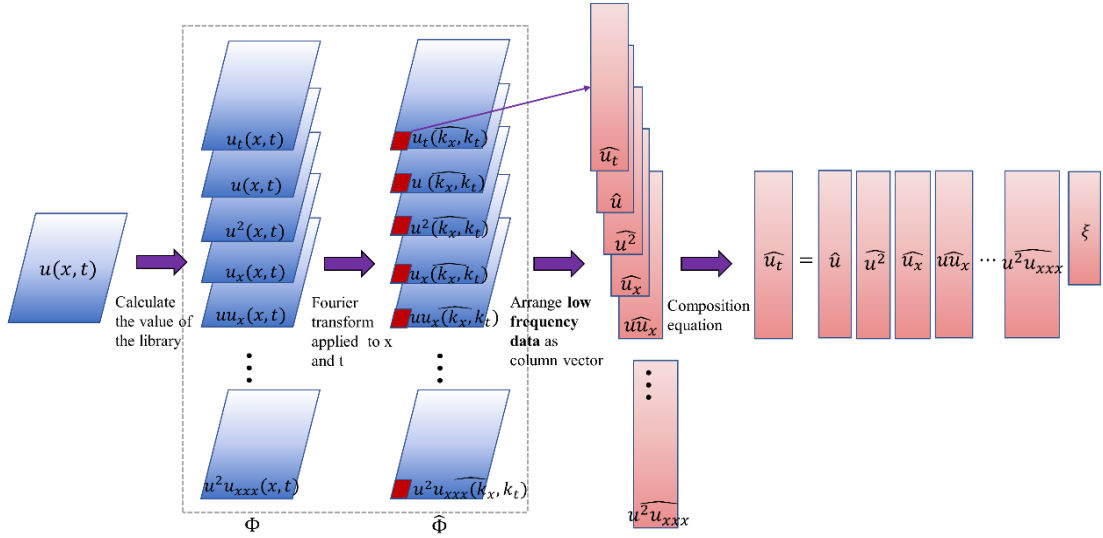

**Fig. 2.** The process of Frequency Domain Identification**.** Obtain the two-dimensional experimental data, calculate the time-space domain value of each term in the library, do two-dimensional Fourier transform to each function value matrix, and arrange the low frequency component of the frequency domain data as column vectors to form the frequency domain matrix equation.

Fig. 3 shows the error cloud pictures of $u_{xxx}$ in time-space domain and frequency domain. We use the polynomial interpolation to calculate the third derivatives $u_{xxx}$ and $un_{xxx}$ of both noiseless data $u(x,t)$ and noise data $un(x,t)$, respectively. And then, the logarithms of their relative errors are calculated in time-space domain, time frequency domain and time-space frequency domain. In Fig. 3, the error of the low frequency component in time-space frequency domain (Fig. 3d) is smaller than the low frequency component in time-frequency domain (Fig. 3c), and is much smaller than the average error in time-space domain (Fig. 3b). This suggests that the error of low frequency component of frequency domain data is small, and the effect of two-dimensional Fourier transform is better than that of one-dimensional Fourier transform. Therefore, using the low frequency data in frequency domain to identify the frequency domain PDE will achieve better results.

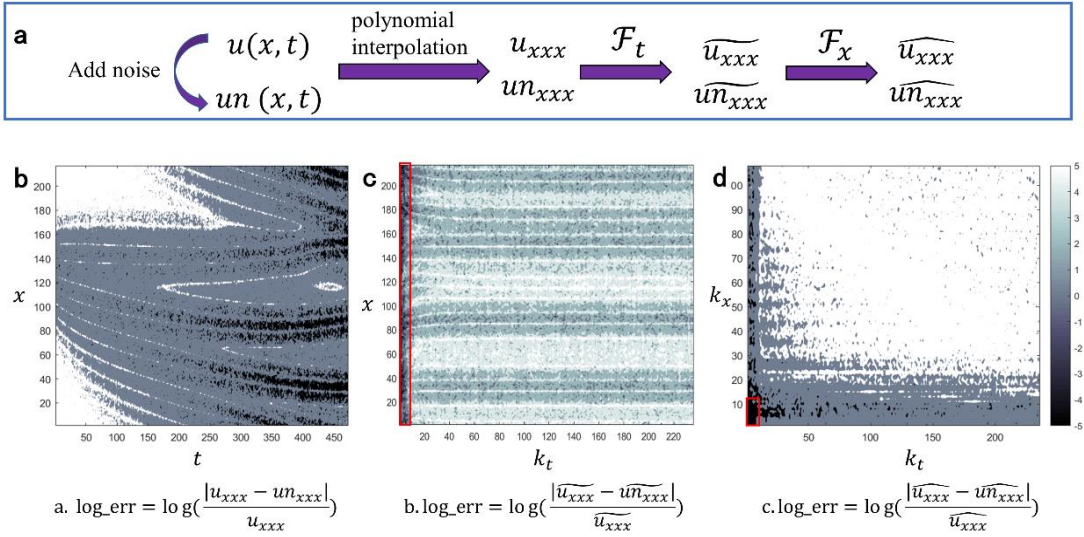

**Fig. 3.** The error cloud pictures in time-space domain and frequency domain. The darker the color, the smaller the error. (a) Process of calculating variables. (b) Logarithm of relative error of $u_{xxx}$ and $un_{xxx}$. (c) Logarithm of relative error of $\widetilde{u_{xxx}}$ and $\widetilde{un_{xxx}}$. (d) Logarithm of relative error of $\widehat{\widetilde{u_{xxx}}}$ and $\widehat{\widetilde{un_{xxx}}}$.

2.3 A new sparse identification method

After obtaining the matrix equation (Eq. 3), we can get sparse vector ξ by solving it. Existing methods [14,17] usually get ξ through the least absolute shrinkage and selection operator (LASSO) [17,27] or the sequential threshold ridge regression [14]. In essence, these methods are based on the least square solution of the matrix equation (Eq. 7), and it is difficult to identify the terms with small coefficients in the PDE. We propose a new idea of sparse identification: true terms (terms in PDE) and false terms (terms not in PDE) have different significances to matrix equation. When the false term is deleted, the matrix equation is still hold; on the contrary, when the true term is deleted, it is no longer hold. Therefore, by detecting the change of the matrix equation after deleting a term in the library, we can determine whether the deleted term is a term in PDE. Our specific method is as follows.

$$\xi = \underset{\xi}{argmin} \ \| U_t - \Phi\xi \|_2^2 + \lambda\|\xi\|_p \qquad (7)$$

First, we calculate the least square solution $\xi_0$ of the matrix equation, and then delete each column $\widehat{\Phi_i}$ of the matrix $\widehat{\Phi}$ and calculate the corresponding least square solutions $\xi_1, \xi_2, \xi_3, \cdots, \xi_n$, respectively. (Eq. 8, where $\widehat{\Phi}^{-1}$ denotes the pseudo inverse of $\widehat{\Phi}$). Finally, we determine whether $\varphi_i$ is in PDE by comparing the changes of $\xi_i$ relative to $\xi_0$. The principle of the method is that the solution of the matrix equation should be completely dependent on the true term and independent of the false term (the coefficient of the false term should be 0). Therefore, the solution of the matrix equation will change greatly when the true term is deleted, and will hardly change when the false term is deleted. After identifying the terms in PDE by the method, we can get the PDE by the least square method.

$$\begin{cases} \xi_0 \colon [\xi_{01}\ \xi_{02}\ \cdots\ \xi_{0n}] = [\widehat{\Phi_1}\ \widehat{\Phi_2}\ \cdots\ \widehat{\Phi_n}]^{-1}\widehat{U_t} \\ \xi_1 \colon [\quad\ \xi_{12}\ \cdots\ \xi_{1n}] = [\quad\ \widehat{\Phi_2}\ \cdots\ \widehat{\Phi_n}]^{-1}\widehat{U_t} \\ \xi_2 \colon [\xi_{21}\quad\ \cdots\ \xi_{2n}] = [\widehat{\Phi_1}\quad\ \cdots\ \widehat{\Phi_n}]^{-1}\widehat{U_t} \\ \quad\quad\quad\quad\quad\vdots \\ \xi_n \colon [\xi_{n1}\ \xi_{n2}\ \cdots\quad\ ] = [\widehat{\Phi_1}\ \widehat{\Phi_2}\ \cdots\quad\ ]^{-1}\widehat{U_t} \end{cases} \qquad (8)$$

We denote the change of the least square solution after $\widehat{\Phi_i}$ is deleted by $q_i$:

$$q_i = \sum_{j=1}^{n} |\xi_{0j} - \xi_{ij}| \quad (i \neq j) \qquad (9)$$

Since we only focus on the relative value of $q_i$, $q_i$ is normalized as $Q_i$ to facilitate comparison.

$$Q_i = \frac{q_i}{\sum_{i=1}^{n} q_i} \qquad (10)$$

$Q_i$ is defined as Candidate Support Rate. We think that the term with a larger Q is the term in PDE, and then the PDE can be obtained by the least square method. Compared with using the sparse regression (Eq. 7) to identify the structure of PDE, the sparse identification method based on Candidate Support Rate (SI-CSR) has the following advantages:

1). The method is simple and effective, and there are no empirical parameters, such as regularization parameter $\lambda$.

2). It can accurately identify the small coefficient terms of the PDE.

3). Equation structure identification and parameters identification are carried out step by step, and the structure of PDE can still be obtained when the parameters identification error is very large.

The above advantages will be demonstrated one by one in Results.

## 3 Results and discussion

In this part, we first show the identification process and results of the Burgers' equation in detail, and then illustrate the effectiveness of the method in this paper for low SNR data through identification results of typical PDEs. Finally, we use the example data of the Burgers' equation to compare this method with other methods to demonstrate its advantages more clearly. (All the example codes in this paper can be downloaded from the website: *https://github.com/Cao-WenBo/PDE_identification*).

We adopt the definition of noise given by Rudy [14]:

$$u_n = u + \alpha \times std(u) \times randn \tag{11}$$

Where $u$ is a clean numerical solution; $\alpha$ is the noise level; $std(u)$ is the standard deviation of $u$; and $randn$ is a random variable the probability distribution of which is a standard normal distribution; $u_n$ represents the data with noise level $\alpha$.

In the following example, for one-, two-, and three-dimensional PDEs, the libraries we build are:

$$\varphi_{1-D} = [u, u^2, u^3, u_x, uu_x, u^2u_x, u^3u_x, u_{xx}, uu_{xx}, u^2u_{xx}, u^3u_{xx}, u_{xxx}, uu_{xxx}, u^2u_{xxx},$$
$$u^3u_{xxx}, u_{xxxx}, uu_{xxxx}, u^2u_{xxxx}, u^3u_{xxxx}]$$

$$\varphi_{2-D} = [u, u^2, u_x, uu_x, u^2u_x, u_{xx}, uu_{xx}, u^2u_{xx}, u_{xxx}, uu_{xxx}, u^2u_{xxx}, u_y, uu_y, u^2u_y,$$
$$u_{yy}, uu_{yy}, u^2u_{yy}, u_{yyy}, uu_{yyy}, u^2u_{yyy}]$$

$$\varphi_{3-D} = [u, u^2, u_x, uu_x, u^2u_x, u_{xx}, uu_{xx}, u^2u_{xx}, u_y, uu_y, u^2u_y, u_{yy}, uu_{yy}, u^2u_{yy}, u_z,$$
$$uu_z, u^2u_z, u_{zz}, uu_{zz}, u^2u_{zz}]$$

3.1 Burgers' equation identification

We first calculate a numerical solution of the Burgers' equation ($u_t = -uu_x + 0.05u_{xx}$), and then add 0%, 10% and 100% noise defined by the Eq. 11 to the numerical solution, respectively. Finally, we calculate the Q corresponding to each term in the library.

**a**

| $\widehat{\varphi}$ | $\hat{u}$ | $\widehat{u^2}$ | $\widehat{u^3}$ | $\widehat{u_x}$ | $\widehat{uu_x}$ | $\widehat{u^2u_x}$ | $\widehat{u^3u_x}$ | $\widehat{u_{xx}}$ | $\widehat{uu_{xx}}$ | $\widehat{u^2u_{xx}}$ |
|---|---|---|---|---|---|---|---|---|---|---|
| Q(0%noise) | 0.0000 | 0.0000 | 0.0000 | 0.0000 | **0.7071** | 0.0000 | 0.0000 | **0.2924** | 0.0000 | 0.0000 |
| Q(10%noise) | 0.0002 | 0.0009 | 0.0007 | 0.0110 | **0.6346** | 0.0066 | 0.0020 | **0.2689** | 0.0090 | 0.0007 |
| Q(100%noise) | 0.0069 | 0.0053 | 0.0033 | 0.0052 | **0.3287** | 0.0155 | 0.0384 | **0.2229** | 0.0246 | 0.0626 |

| $\widehat{\varphi}$ | $\widehat{u^3u_{xx}}$ | $\widehat{u_{xxx}}$ | $\widehat{uu_{xxx}}$ | $\widehat{u^2u_{xxx}}$ | $\widehat{u^3u_{xxx}}$ | $\widehat{u_{xxxx}}$ | $\widehat{uu_{xxxx}}$ | $\widehat{u^2u_{xxxx}}$ | $\widehat{u^3u_{xxxx}}$ |
|---|---|---|---|---|---|---|---|---|---|
| Q(0%noise) | 0.0000 | 0.0000 | 0.0000 | 0.0001 | 0.0000 | 0.0000 | 0.0000 | 0.0000 | 0.0000 |
| Q(10%noise) | 0.0056 | 0.0152 | 0.0042 | 0.0138 | 0.0059 | 0.0005 | 0.0125 | 0.0005 | 0.0070 |
| Q(100%noise) | 0.0287 | 0.0094 | 0.0146 | 0.0320 | 0.0488 | 0.0255 | 0.0211 | 0.0698 | 0.0369 |

**b**

| | Correct PDE | |
|---|---|---|
| | Correct PDE | $u_t = -uu_x + 0.05u_{xx}$ |
| | Identified PDE(0%noise) | $u_t = -1.0000uu_x + 0.0500u_{xx}$ |
| | Identified PDE(10%noise) | $u_t = -0.9981uu_x + 0.0504u_{xx}$ |
| | Identified PDE(100%noise) | $u_t = -0.9429uu_x + 0.0498u_{xx}$ |

**c** (plot of exact data vs 100% noise data)

**Fig. 4.** Identification results of Burgers' equation. When α = 0, the numerical differential is calculated by difference. When α ≠ 0, it is calculated by polynomial interpolation. (a) The Q of each term in the library with different noise levels. (b) Parameters identification results of different noise levels. (c) Data with 100% noise.

Fig. 4a shows that when the noise level α = 0, the Q of $\widehat{uu_x}$ and $\widehat{u_{xx}}$ are much larger than other terms. Therefore, we consider that the corresponding time-space domain forms $uu_x$ and $u_{xx}$ are the terms in PDE. At the same time, it is noted that the Q of $\widehat{uu_x}$ and $\widehat{u_{xx}}$ are still obviously larger than other terms when the data has 10% or even 100% noise, that is to say, even if the data contains 100% noise (Fig. 4c), we can still identify the structure of PDE from the noise data, which proves the robustness of the method to noise. After determining that the terms $uu_x$ and $u_{xx}$ are in PDE, the least square solution is carried out to obtain the parameters of the PDE (Fig. 4b). It can be seen that when $\alpha < 10\%$, the parameters identification error is less than 1%. Even if $\alpha = 100\%$, the maximum error of the parameters is only 6%, which further illustrates the robustness of Frequency Domain Identification to noise.

**a**

| PDE | Form | $\alpha_{max}$ | Mean Relative Error(MRE) | | |
|---|---|---|---|---|---|
| | | | α = 0% | α = 5% | α = 100% |
| Burgers | $u_t = -uu_x + 0.05u_{xx}$ | 200% | 0.02% | 0.37% | 5.32% |
| KS | $u_t = -uu_x - 0.7u_{xx} - u_{xxx} - 1.3u_{xxxx}$ | 175% | 0.01% | 1.78% | 3.96% |
| KdV | $u_t = -0.5u_x - 1.5uu_x - 0.25u_{xxx}$ | 160% | 0.02% | 0.66% | 16.6% |

**b**

| PDE | Form | $\alpha_{max}$ | Mean Relative Error(MRE) | | |
|---|---|---|---|---|---|
| | | | α = 0% | α = 5% | α = 100% |
| 2-D Wave | $u_{tt} = u_{xx} + u_{yy}$ | 105% | 0.14% | 0.33% | 7.34% |
| 2-D Burgers | $u_t = -uu_x + 0.01u_{xx} - uu_y + 0.01u_{yy}$ | 185% | 0.07% | 0.13% | 2.98% |

**c**

| PDE | Form | $\alpha_{max}$ | Mean Relative Error(MRE) | | |
|---|---|---|---|---|---|
| | | | α = 0% | α = 250% | α = 1000% |
| 3-D Wave | $u_t = u_{xx} + 1.5u_{yy} + 2u_{zz}$ | 1500% | 3.10% | 84.2% | 635% |
| 3-D Burgers | $u_t = -uu_x + 0.1u_{xx} - uu_y + 0.1u_{yy} - uu_z + 0.1u_{zz}$ | 850% | 5.54% | 4.86% | 16.3% |

**Table. 1.** Summary of identification results of typical equations. $\alpha_{max}$ represents the maximum allowable noise, and when α < $\alpha_{max}$, we can correctly identify the structure of PDE. MRE represents the average relative error of all parameters in the PDE. (a) Results of 1-D PDE identification. All the results are the average values of 10 groups of different random noise. (b) Results of 2-D PDE identification. All the results are the average values of 10 groups of different random noise. (c) Results of 3-D PDE identification. All the results are the average values of 5 groups of different random noise (limited to the amount of calculation).

## 3.2 One-dimensional PDE identification

Table. 1a shows the identification results of the Burgers, Korteweg–de Vries (KdV) and Kuramoto-Sivashinsky (KS) equations. We can see that the corresponding $\alpha_{\max}$ of the three PDEs are all greater than 150%, which indicates that when $\alpha < 150\%$, the method can accurately identify the structure of PDE, in contrast, Rudy [14] identified the PDEs from only 1% noise data. For the results of parameters identification, when $\alpha = 5\%$, the MRE of the three PDEs is less than 2%, and even if $\alpha = 100\%$, the MRE of the Burgers' equation and the KS equation is less than 6%. The error is much less than that those of the time-space domain methods [14,17]. Two methods will be compared in detail in the next section. It is worth mentioning that the KS and KdV equations contain high-order derivative terms such as $u_{xxx}$ and $u_{xxxx}$, which is a challenge to the existing identification methods, because the higher the derivative order, the greater the numerical differential error for the noise data. As for the method in this paper, the high-order derivative terms in PDE can be easily identified from low SNR data, which reflects its advantages.

## 3.3 Two-dimensional PDE identification

Table. 1b shows the identification results of 2D Wave and 2D Burgers' equations to show the effectiveness of the method for two-dimensional PDE. Different from using the polynomial interpolation [28] to calculate the numerical differentiation for one-dimensional PDE, for two-dimensional or three-dimensional PDE, that is, three-dimensional or four-dimensional data, we only use difference to calculate the numerical differentiation due to the large amount of computation of polynomial interpolation. Table. 1b shows that the method can obtain good results for two-dimensional PDE even if the numerical differentiation is calculated by difference. In fact, this is because we can do Fourier transform along three dimensions for two-dimensional PDE (one more than one-dimensional PDE), which can eliminate the influence of noise to a greater extent, and this viewpoint will be explained in detail in the identification of three-dimensional PDE. In addition, it is noted that the coefficients of $u_{xx}$ and $u_{yy}$ in 2D Burgers' equation are only 1% of those of other terms, which is quite challenging for the existing method based on the least square solution of the matrix equation, but the SI-CSR can still identify the correct structure of PDE, which reflects the advantage of the method in identifying terms with smaller coefficients in the PDE.

3.4 Third-dimensional PDE identification

The core of Frequency Domain Identification is to filter noise through Fourier transform. Fig. 3 shows that the more times of Fourier transform, the smaller the error of low frequency component. Therefore, for high-dimensional PDEs, because the Fourier transform can be carried out along more dimensions, better results can be achived. Table. 1c shows that the method can accurately identify the structure of three-dimensional PDE from very low SNR data. But it is also easy to notice that the error of parameters identification is very large even for noiseless data, because we use a larger step size to filter high-frequency noise in the difference process. This leads to a large truncation error. Usually the derivative is calculated by the polynomial interpolation and then identified in frequency domain, and the parameters error will be small, but it is expensive for three-dimensional PDE, so we do not use the polynomial interpolation to calculate the derivative. In addition, the larger parameters identification error also reflects another advantage of the SI-CSR: the processes of PDE structure identification and parameters identification are carried out step by step, which can accurately identify the structure of PDE when the parameters identification error is large.

3.5 Compare Frequency Domain Identification with other methods

In order to further verify the advantage of Frequency Domain Identification to deal with noise, we take the Burgers' equation as an example and assume that the structure of the PDE is known. Then we compare the errors of the PDE parameters obtained by the time-space domain identification methods [14,17], by Frequency Domain Identification and the time-space domain identification after the low pass filtering to confirm the PDE identification accuracy of Frequency Domain Identification for noise data.

As mentioned earlier, the time-space domain identification method is to get PDE parameters by solving the time-space domain equation $[\xi_1, \xi_2] = [UU_x, U_{xx}]^{-1} U_t$, while Frequency Domain Identification gets PDE parameters by solving the frequency domain equation $[\xi_1, \xi_2] = [\widehat{UU_x}, \widehat{U_{xx}}]^{-1} \widehat{U_t}$. In the low pass filtering method, $u$ is filtered at first, and then PDE parameters are calculated by using the time-space domain identification method. The parameters are calculated by different methods for different noise levels $\alpha$, and the average of results of 10 groups of random noise is taken to get Fig. 5. It can be seen that for different noise levels $\alpha$, the error of PDE parameters obtained by Frequency Domain Identification is much smaller than that of the time-space domain identification method, and also significantly smaller than that of the low pass filtering method, especially for strong noise data. At the same time, we can also see the results of low pass filtering with different thresholds. When use low pass filtering with a low threshold, the signal will be distorted but more noise can be filtered; while with a high threshold, it is on the contrary. Therefore, when α is small, the error of the former is large because it makes the signal distorted; while when α is large, it is small because it filters more noise. This shows that it is difficult for low pass filtering to balance the contradiction between signal fidelity and noise reduction, that is, filtering noise is at the cost of signal distortion.

The very accurate parameters identification results of Frequency Domain Identification also show its potential in equation parameters identification, such as the need to identify the damping of the system when the basic form of the structural motion equation is known. It is possible that Frequency Domain Identification can obtain more accurate results than other methods.

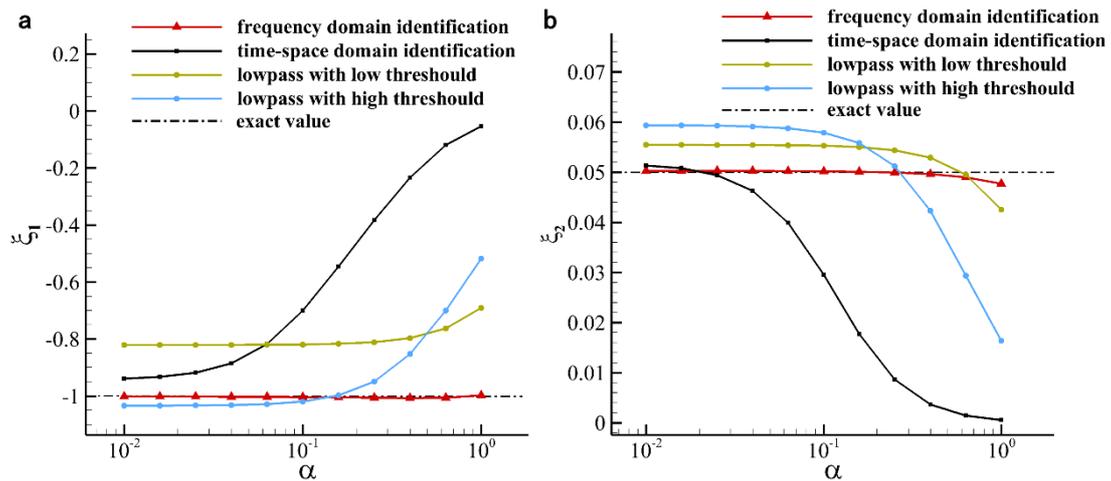

**Fig. 5.** Parameters identification results of different methods under different noise levels. (a) Coefficient of $uu_x$. (b) Coefficient of $u_{xx}$.

3.6 Compare SI-CSR with STLM

To solve the sparse vector ξ of the matrix equation (Eq. 3), the existing methods [5,14,17] are essentially based on the least square solution, which are easy to ignore the terms with small coefficients (such as $u_{xx}$ in the Burgers' equation). In this paper, we first calculate the frequency domain matrix equation (Eq. 6) of the example data of the Burgers' equation, and then use the sequential threshold least square method (STLM) [5] and the SI-CSR to solve the vector ξ.

For convenience, the two methods are simplified to compare their effects. For the STLM, the term with the smallest coefficient of least square solution of the matrix equation is deleted until there are only two remaining terms. For the SI-CSR, only the two largest terms of Q are taken. We compare the effects of the two methods by comparing whether the correct partial differential terms are identified by them.

A total of 10 groups of random noise are used to calculate the average robustness of the two methods. The results show that for the same matrix equation, the SI-CSR can identify the Burgers equation from 200% noise data, while the sequential threshold least square method can only identify it from 12% noise data. For larger noise, it cannot identify $u_{xx}$. And for 2-D Burgers' equation, even if there is no noise in the data, STLM cannot correctly identify the structure of the equation. This indicates the advantage of the SI-CSR to identify the terms with small coefficients in the equation.

## 4  Conclusions

In this paper, we first propose a general noise filtering framework (Eq. 4) for the PDE identification process, for any linear noise filtering method, it can filter the noise while ensuring that the data still obey the original partial differential equation. Then, we propose Frequency Domain Identification based on low-pass filtering and apply it to the framework. This method eliminates the influence of noise to the greatest extent, and has high accuracy and robustness for the identification of partial differential equations with low signal-to-noise ratio data. In many examples, this method can identify equations from data with 100% noise (compared with 1% of Ref. [14]), which greatly reduces the accuracy requirements for measurement data and makes it possible to find partial differential equations from experimental data with high noise. At the same time, this method can also be applied to other system identification problems. In addition, SI-CSR is a simple and reliable method to identify the structure of PDE, in the method, the processes of PDE structure identification and parameters identification are carried out step by step, which can accurately identify the structure of PDE even when the equation fitting error is large. Moreover, the identification results of two-dimensional and three-dimensional equations also show that our method can effectively deal with high-dimensional measurement data, and it can identify the three-dimensional PDE structure from the data with 850% noise in our examples, it is very amazing.

It is worth noting that although we identify the equation through the low frequency component of the frequency domain data, we do not assume that the experimental data must be low frequency, as long as there are enough low frequency components in the data. Even if the experimental data are almost completely high-frequency in one dimension, Fourier transform can be performed in other dimensions and the low frequency components of them can be selected to identify PDE to eliminate noise as much as possible. We will explain this in our future work.

**Declarations**

**Funding**

No funds, grants, or other support was received.

**Conflicts of interest/Competing interests**

The authors have no conflicts of interest to declare that are relevant to the content of this article.

**Availability of data and material**

The datasets generated and analysed during the current study are available in the *https://github.com/Cao-WenBo/PDE_identification*.

**Code availability**

All the example codes in this paper can be downloaded from the website: *https://github.com/Cao-WenBo/PDE_identification.*

**Ethics approval**

Not applicable

**Consent to participate**

All authors contributed to the study conception and design and commented on previous versions of the manuscript.

**Consent for publication**

All the authors agreed to the publication of the manuscript.